\UseRawInputEncoding
\documentclass[twocolumn,           % Format : preprint, twocolumn
               showpacs,            % Pacs : showpacs, noshowpacs
               nopreprintnumbers,     % Preprint: preprintnumbers,
               aps,                 % Society: ...
               prd,          	    % Journal Style : pra, prb, prc, prd, pre,
               letterpaper,             % Size : a4paper, ...
              groupeaddress,      % Affiliation (Title) : groupedaddress,
               nofootinbib,         % Footnote: footinbib, nofootinbib
               tightenlines,        % Remove additional spaces in a line
               floats,floatfix,      % Floating pictures and tables
               showkeys
               ]{revtex4-1}
               
\usepackage[toc,page]{appendix}
\usepackage{graphicx}% Include figure files
\usepackage{dcolumn}% Align table columns on decimal point
\usepackage{bm}% bold math
\usepackage{amsmath}
\usepackage{amsfonts,amssymb}
\usepackage{soul}
\usepackage{color}
\definecolor{v}{rgb}{0.6, 0.2, 0.8} %comentarios VM
\definecolor{MAGA}{rgb}{0.1, 0.43, 0.75}
\definecolor{jm}{rgb}{0.13, 0.48, 0.64}

\usepackage{xcolor}
\usepackage{enumerate}
\usepackage{float}
\usepackage{subfigure}
\usepackage{multirow,tabularx, booktabs}

\usepackage{orcidlink}

\usepackage{silence}
\begin{document}

\title{Phenomenological emergent dark energy in the light of DESI Data Release 1}

\newcommand{\orcidauthorA}{0000-0003-0405-9344} % Alberto
\newcommand{\orcidauthorD}{0000-0002-4276-2906} % Luisa
\newcommand{\orcidauthorB}{0000-0002-6356-8870} % Miguel
\newcommand{\orcidauthorC}{0000-0003-4446-7465} % Veronica

\author{A.  Hern\'andez-Almada$^1$\orcidlink{\orcidauthorA}}
\email{ahalmada@uaq.mx, corresponding author}

\author{M. L. Mendoza-Mart\'inez$^{2}$\orcidlink{\orcidauthorD}}
\email{ml.mendoza@tec.mx}

\author{Miguel A. Garc\'ia-Aspeitia$^3$\orcidlink{\orcidauthorB}}
\email{angel.garcia@ibero.mx}

\author{V. Motta$^4$\orcidlink{\orcidauthorC}}
\email{veronica.motta@uv.cl}

\affiliation{$^1$ Facultad de Ingenier\'ia, Universidad Aut\'onoma de
Quer\'etaro, Centro Universitario Cerro de las Campanas, 76010, Santiago de 
Quer\'etaro, M\'exico}
\affiliation{$^2$ Tecnologico de Monterrey, Escuela de Ingenier\'ia y Ciencias, Epigmenio Gonz\'alez 500, San Pablo, 76130, Santiago de Quer\'etaro Qro. M\'exico.}
\affiliation{$^3$ Depto. de F\'isica y Matem\'aticas, Universidad Iberoamericana Ciudad de M\'exico, Prolongaci\'on Paseo \\ de la Reforma 880, M\'exico D. F. 01219, M\'exico}
\affiliation{$^4$Instituto de F\'isica y Astronom\'ia, Universidad de Valpara\'iso, Avda. Gran Breta\~na 1111, Valpara\'iso, Chile.}

%-------------------------------------------------------------------------------------------------
%-------------------------------------------------------------------------------------------------
\begin{abstract}
This manuscript revisits the phenomenological emergent dark energy model (PEDE) by confronting it with recent cosmological data from early and late times. In particular we analyze PEDE model by using the baryon acoustic oscillation (BAO) measurements coming from both Dark Energy Spectroscopy Instrument (DESI) data release 1  and Sloan Digital Sky Survey (SDSS). Additionally, the measurements from cosmic chronometers, supernovae type Ia (Pantheon+), quasars, hydrogen II galaxies and cosmic background radiation distance priors are considered. By performing a Bayesian analysis based on Monte Carlo Markov Chain, we find consistent results on the constraints when SDSS and DESI are considered. However, we find higher values on the Hubble constant than Supernova $H_0$ for the Equation of State (SH0ES) does although it is still in agreement, within $1\sigma$ confidence level, when BAO measurements are added. Furthermore, we estimate the age of the Universe younger $\sim3\%$ than the one predicted by the standard cosmology. Additionally, we report  values of $q_0 = -0.771^{+0.007}_{-0.007}$, $z_T = 0.764^{+0.011}_{-0.011}$ for the deceleration parameter today and the deceleration-acceleration transition redshift, respectively.
However, PEDE cosmology is disfavoured by the combined samples.
\end{abstract} 
%\draft
\pacs{Dark energy, PEDE, cosmology}
%\date{=day}
\maketitle

%%%%%%%%%%%%%%%%%%%%%%%%%%%%%%%%%%%%
\section{Introduction}
%%%%%%%%%%%%%%%%%%%%%%%%%%%%%%%%%%%%

The description of our Universe is one of the most challenging topics in modern physics. This is because there are old unsolved problems, like the existence of dark matter (DM) related to the evolution of the structure in our Universe and the observed current acceleration \cite{Riess:1998,Perlmutter:1999}, described by a de Sitter geometry and possibly caused by a dark energy (DE) fluid.
In the DM case, we have many candidates, from supersymmetric particles to scalar fields \cite{Arbey:2021gdg}, while for DE we have modifications to Einstein Field equations to different fluids that mimics the accelerated evolution \cite{Motta:2021hvl,Herrera-Zamorano:2020rdh,Hernandez-Almada:2018osh}. However, the best candidates are a DM that behaves as cold fluid and a DE as the well known cosmological constant. Is in this vein, that the $\Lambda$-Cold Dark Matter ($\Lambda$CDM) emerges, having unprecedented achievements in the understanding of the Universe evolution.
Despite the success of the $\Lambda$CDM model, we do not have a full understanding of, for example, what the cosmological constant is. In this case, our best calculations predict an ultraviolet divergence when all the quantum vacuum fluctuations are calculated through the quantum field theory. This problem is known as the fine tuning problem (see \cite{Weinberg,Zeldovich}).

On the other hand, we have new challenges like those related to the $H_0$ and
$\sigma_8$ tension detected by several observations \cite{DiValentino:2021izs,Motta:2021hvl, Bamba_2012,DiValentino:2020vhf,DiValentino:2020zio,DiValentino:2020vvd,DiValentino:2020srs,Perivolaropoulos:2021jda,Abdalla:2022yfr}. Regarding the $H_0$ tension, this is due to the discrepancy between observations based on the Cosmic Microwave Background Radiations (CMB) by Planck 2018 which reports a value of $H_0 = 67.36 \pm 0.54\, {\rm km/s/Mpc}$ \cite{Planck:2018} and local observations given by Supernovae of the Ia type (SNIa) based in SH0ES measurements with a reported value of $H_0 =  73.04 \pm 1.04\, {\rm km/s/Mpc}$ \cite{Riess:2019}.  The disagreement is approximately at $\sim5\sigma$ and, until now, we do not know if it is caused by systematic errors or it is due to mistaken theoretical assumptions. However, if the tension is confirmed, this event open a new window for possible new physics beyond the standard cosmological model.
In addition, there is also tension in the amplitude of matter perturbations smoothed over $8h^{-1}$Mpc, known as $\sigma_8$, between the $\Lambda$CDM inferred value from CMB anisotropies and those obtained from weak gravitational lensing \cite{DES:2021wwk,KiDS:2020suj, Heymans_2021}.

Between the cosmological models to solve these problems we have quintom cosmology \cite{Yang_2024}, $f(R)$ theories \cite{Pati2024}, extra dimensions \cite{Garcia-Aspeitia:2018fvw, Aspeitia:2008}, fractional calculus \cite{Garcia-Aspeitia:2022uxz}, effective fluids with DE parametrizations \cite{giare2024, calderon2024desi2024reconstructingdark, DESI:2024kob}, the effects of neutrinos in cosmology \cite{Sunny:2024}, entropy models \cite{Mendoza_Martinez_2024, Almada:Kaniadakis:2022, Leon_2021}, interacting dark energy-dark matter models \cite{Pan_2019}, scalar fields \cite{Yin_2022, Yin2024}, bimetric gravity \cite{dwivedi20242dbaovs3d, Marcus:2021} and amount others  \cite{dwivedi20242dbaovs3d}. For some reviews about these alternatives see for instance \cite{Motta:2021hvl, DiValentino:2021izs, hu2023hubbletensionevidencenew, Khalife_2024}.

A natural and interesting alternative to resolve the problem of the Universe acceleration is based on the phenomenological emergent dark energy model (PEDE) proposed by \cite{PEDE:2019ApJ} and later studied in \cite{Pan:2019hac, Hernandez-Almada:2020uyr,Li:2020ybr, Koo:2020ssl, Yang_2021, DESI:2024kob, Liu:2024wne}. In this case, PEDE has the same parameters of $\Lambda$CDM, however, the causative of the Universe acceleration is a function that contains a tangent hyperbolic behavior with an inner logarithmic function. An interesting characteristic  of the PEDE function is that it is negligible at early times,  i.e. not affecting the physics of the early Universe, but at late times the contribution emerge  producing the observed late time acceleration. Li and Shafieloo \cite{PEDE:2019ApJ} constrain the model using SNIa, CMB and baryon acoustic oscillations (BAO),  claiming a possible alleviation to the $H_0$ tension. Later on, the authors \cite{Pan:2019hac} show  such alleviation  within $68\%$ of the confidence level by using mainly CMB observational data. In the same vein, \cite{Koo:2020ssl} reconstruct the cosmic expansion from SNIa data using a non-parametric iterative smoothing method. In \cite{Hernandez-Almada:2020uyr} the authors constrain PEDE using homogeneous and non-homogeneous Observational Hubble Data (OHD) samples. They report values consistent with $\Lambda$CDM and implement a robust dynamical system analysis, presenting the main stability points for this case. 

 Additionally, in \cite{DESI:2024kob} authors constraint different models, specifically PEDE and its extension known as Generalized Dark Energy (GEDE), using DESI 2024 and CMB datasets. Although the values of the constrained parameters exhibit significant variations, it is important to mention that GEDE carries an additional free parameter ($\Delta$) that could be the cause of these differences (for details see \cite{DESI:2024kob}) and \cite{Li:2020ybr}.
Finally, we need to mention a recent paper \cite{Liu:2024wne} where PEDE is studied with a barotropic fluid as a DM, called PEDE$+w_{dm}$. In this case, the authors implement constraints using CMB, BAO, SNIa  with priors on $H_0$  and including recent datasamples \cite{descollaboration2024dark,desicollaboration2024desi,bunker2024jades}. Their results show, for the joint analysis, an increase of $H_0$ tension at $1.5\sigma$  but an alleviation of $S_8$ tension at $0.4\sigma$.  Thus, they conclude that the PEDE$+w_{dm}$  is not a good alternative to the PEDE model.

The outline of the paper is as follows: Section \ref{sec:cosmology} is dedicated to present a revision of the PEDE model and its description under a Friedmann-Lemaitre-Robertson-Walker cosmology. Section \ref{sec:constraints} contains the datasets used to constrain the model. In Section \ref{Results} we present the results of our study, and finally in Section \ref{SD} we  show our results and a discussion.

%%%%%%%%%%%%%%%%%%%%%%%%%%%%%%%%%%%%
\section{Phenomenological emergent dark energy cosmology} \label{sec:cosmology}
%%%%%%%%%%%%%%%%%%%%%%%%%%%%%%%%%%%%

In order to explore the PEDE cosmology, we use the Friedmann-Lemaitre-Robertson-Walker (FLRW) line element with flat curvature $ds^2=-dt^2+a(t)^2(dr^2+r^2d\Omega^2)$, where $a(t)$ is the scale factor and $d\Omega^2$ is the solid angle. In this case, we assume a perfect fluid energy-momentum tensor $T_{\mu\nu}=pg_{\mu\nu}+(\rho+p)u_{\mu}u_{\nu}$, where $p$, $\rho$ and $u_{\mu}$ are the pressure, density and  quadri-velocity respectively, with only time dependence in order to maintain the cosmological principle.

Using the Einstein field equations, PEDE cosmology is described by the dimensionless Friedmann equation, which takes the form

\begin{equation}
    E(z)^2=\Omega_{0m}(z+1)^3+\Omega_{0r}(z+1)^4+\Omega_{DE}(z),
\end{equation}
where $E(z)\equiv H(z)/H_0$, $\Omega_{0m}$, $\Omega_{0r}$ are matter and radiation density parameters respectively, $z$ is the redshift and the mathematical structure of $\Omega_{DE}(z)$  is the DE density parameter associated with PEDE as proposed by Li and Shafieloo \cite{PEDE:2019ApJ}
\begin{equation}
    \Omega_{DE}(z)=\Omega_{0DE}[1-\tanh(\log_{10}(z+1))]. \label{eq:PEDE}
\end{equation}
Notice that PEDE model   has the same free parameters as $\Lambda$CDM cosmology. Thus, the deceleration parameter read as
\begin{eqnarray}
    q(z)&=&\frac{z+1}{2E(z)^2}\Big[3\Omega_{0m}(z+1)^2+4\Omega_{0r}(z+1)^3\nonumber\\
    &&-\Omega_{0DE}\frac{{\rm sech}^2\Big[\frac{\ln(z+1)}{\ln(10)}\Big]}{\ln(10)(z+1)}\Big]-1.
\end{eqnarray}
Additionally, it is possible to write the equation of state (EoS) only caused by the PEDE dynamics, which takes the form

\begin{eqnarray}
    w(z)_{PEDE}=-\frac{1}{3\ln(10)}(1+\tanh[{\rm log}_{10}(z+1)])-1.
\end{eqnarray}
We can write the effective EoS  where all the densities contributes as

\begin{eqnarray}
    w(z)_{eff}&=&\frac{1}{3}\Big[\frac{z+1}{E(z)^2}\Big(3\Omega_{0m}(z+1)^2+4\Omega_{0r}(z+1)^3\nonumber\\
    &&-\Omega_{0DE}\frac{{\rm sech}^2\Big[\frac{\ln(z+1)}{\ln(10)}\Big]}{\ln(10)(z+1)}\Big)-3\Big]\,,
\end{eqnarray}
calculated through the formula $3w(z)_{eff}=2q(z)-1$ where $q(z)$ is the deceleration parameter.

%%%%%%%%%%%%%%%%%%%%%%%%%%%
\section{Datasets and constraints} \label{sec:constraints}
%%%%%%%%%%%%%%%%%%%%%%%%%%%

The PEDE parameter phase-space is given by $\Theta=\{h, \Omega_{0b}, \Omega_{0m}\}$, where $h$ is the dimensionless Hubble constant, and $\Omega_{0b}$ and $\Omega_{0m}$ are the baryon and matter densities, respectively. The aim is to constrain them using CMB distance priors \cite{Planck:2018}, cosmic chronometers (CC) \cite{Moresco:2016mzx, Jiao_2023, Tomasetti_2023}, SNIa \cite{Scolnic2018-qf, Brout_2022}, QSO \cite{ShuoQSO:2017}, hydrogen II galaxies (HIIG) \cite{GonzalezMoran2019, Gonzalez-Moran:2021drc} and BAO datasets by applying Monte Carlo Markov Chain (MCMC) tools through the Emcee package \cite{Foreman:2013} under Python environment. Hence MCMC runs are stopped up to achieve the convergence of the chains based on  the autocorrelation function. Additionally, the priors considered are Uniform distributions over $h:[0.2,1]$ and $\Omega_{0m}:[0,1]$, and a Gaussian distribution centered at $\Omega_{0b}h^2 = 0.02202\pm 0.00046$ \cite{Planck:2018}. We define a baseline dataset as the combination of the CMB distance prior, CC, SNIa, QSO, HIIG datasets. Hence, the figure-of-merit for the baseline analysis is obtained by building a Gaussian log-likelihood given as
\begin{equation} \label{eq:chi2_joint}
-2\ln(\mathcal{L}_{\rm baseline})\varpropto \chi^2_{\rm baseline} = \sum_i \chi^2_i \,,
\end{equation}
where the sum runs over each $\chi^2$ according to the baseline dataset. Additionally, we add BAO data to this baseline.  In the following, a brief description of each sample is presented. 

%%%%%%%%%%%%%%%%%%%%%%%%%%%%%%%%%%%%%
\subsection{Cosmic Microwave Background}
%%%%%%%%%%%%%%%%%%%%%%%%%%%%%%%%%%%%%%

The most recent CMB anisotropy data come from Planck 2018 legacy dataset release \cite{Planck:2018} and provide us strong constraints on the cosmological model parameters. A way to extract such constraints without using a full perturbative analysis is available through the named distance prior, which  gives effective information from the CMB power spectrum through the acoustic scale $l_A$, the shift parameter $R$ and $\Omega_{0b}h^2$. The parameter $l_A$ is related to the CMB temperature power spectrum in the transverse direction  and generates variations of the peak spacing, $R$ characterizes the 
temperature power spectrum along the line-of-sight  direction and modifies the heights of the peaks.

The figure-of-merit is built as
\begin{equation}
    \chi^2_{\rm CMB} = \vec{X}^T {\rm Cov}_{\rm CMB}^{-1} \vec{X}\,,
\end{equation}
where $\vec{X}$ is the difference vector between the theoretical and observed quantities 
\begin{align}
\vec{X} &= \begin{pmatrix}
           R -  1.7493           \\
           l_A - 301.462         \\
           \Omega_{0b} - 0.02239
         \end{pmatrix}
\end{align}
and $\rm{Cov}_{\rm CMB}^{-1}$ is the inverse of the covariance matrix given by
\begin{align}
{\rm Cov}_{\rm CMB}^{-1} &= \begin{pmatrix}
            93.2052939 &   -1.34477655  &   1634.16998    \\
           -1.34477655 &    0.160560491 &  5.05824338     \\
           1634.16998  &    5.05824338  &  78905.7112
         \end{pmatrix} \times 10^{3},
\end{align}
estimated using the CMB distance priors derived by \cite{Chen_2019} based on the finally released Planck TT, TE, EE+lowE data in 2018 at $l\geq 30$ and $\omega$CDM model. The theoretical value of $R$ is estimated by
\begin{equation}
    R = \sqrt{\Omega_{0m}H_0^2}\, r(z^*)\,,
\end{equation}
where $r(z^*)$ is the comoving distance from the observer to the redshift  at the photon decoupling epoch $z^*$ given by
\begin{equation}
    r(z^*) = \frac{c}{H_0}\int_0^{z^*} \frac{dz'}{E(z')}\, ,
\end{equation}
 where $c$ is the speed of light.

%%%%%%%%%%%%%%%%%%%%%%%%%%%%%%%%%%%%%
\subsection{Cosmic Chronometers}\label{sec:cc}
%%%%%%%%%%%%%%%%%%%%%%%%%%%%%%%%%%%%%%

Cosmic chronometers  is the name of a useful cosmological model independent dataset of 33 Hubble parameter measurements  obtained from differential age tools applied to galaxies \cite{Moresco:2016mzx, Jiao_2023, Tomasetti_2023}. These points are considered uncorrelated and cover a redshift $0.07<z<1.965$, thus a $\chi^2$-function can be used in the form
\begin{equation} \label{eq:chi2_OHD}
    \chi^2_{{\rm CC}}=\sum_{i=1}^{33}\left(\frac{H_{th}(z_i, \Theta)-H_{obs}^i}{\sigma^i_{obs}}\right)^2,
\end{equation}
where the sum runs over each point in the dataset, $H_{th}(z_i)$ is the theoretical Hubble parameter at the redshift $z_i$ and $(H_{obs}) \pm \sigma_{obs}$ is the observational counterpart and its  uncertainty,  respectively.

%%%%%%%%%%%%%%%%%%%%%%%%%%%%%%%%%%%%%%
\subsection{Type Ia Supernovae}
%%%%%%%%%%%%%%%%%%%%%%%%%%%%%%%%%%%%%%

The Pantheon+ dataset \cite{Scolnic2018-qf, Brout_2022} is the largest sample of SNIa with a total of 1701 correlated distance modulus measurements which cover a redshift range of $0.001<z<2.26$. The theoretical distance modulus is defined by 
\begin{equation}\label{eq:mu_sn}
    m_{th}=\mathcal{M}+5\log_{10}\left[\frac{d_L(z)}{10\, pc}\right],
\end{equation}
where $\mathcal{M}$ is a nuisance parameter and $d_L(z)$ is the luminosity distance estimated as
\begin{equation}\label{eq:dL}
    d_L(z)=(1+z)\frac{c}{H_0}\int_0^z\frac{dz^{\prime}}{E(z^{\prime})}.
\end{equation}
%where $c$ is the speed of light.
To avoid including this nuisance parameter in the statistics inference, it is convenient to use the following $\chi^2$ function \cite{Conley2010}
\begin{equation}\label{eq:chi2SnIa}
    \chi_{\rm SNIa}^{2}=a +\log \left( \frac{e}{2\pi} \right)-\frac{b^{2}}{e},
\end{equation}
where
\begin{eqnarray}
    a &=& \Delta\boldsymbol{\tilde{\mu}}^{T}\cdot\mathbf{Cov_{P}^{-1}}\cdot\Delta\boldsymbol{\tilde{\mu}}, \nonumber\\
    b &=& \Delta\boldsymbol{\tilde{\mu}}^{T}\cdot\mathbf{Cov_{P}^{-1}}\cdot\Delta\mathbf{1}, \\
    e &=& \Delta\mathbf{1}^{T}\cdot\mathbf{Cov_{P}^{-1}}\cdot\Delta\mathbf{1}\, , \nonumber
\end{eqnarray}
and $\Delta\boldsymbol{\tilde{\mu}}$ is the vector of the difference between the theoretical and observed distance modulus, $\Delta\mathbf{1}=(1,1,\dots,1)^T$ is
the transpose of the unit vector
and $\mathbf{Cov_{P}}$ is the covariance matrix.

%%%%%%%%%%%%%%%%%%%%%%%%%%%%%%%%%%%%%
\subsection{Quasars}
%%%%%%%%%%%%%%%%%%%%%%%%%%%%%%%%%%%%%%

A sample of 120 distance modulus measurements of intermediate-luminosity quasars (QSO) spanning a redshift range $0.462<z<2.73$ is presented in \cite{ShuoQSO:2017}. We constrain the cosmological parameters through the $\chi^2$ function
\begin{equation}
    \chi^2_{\rm QSO} = \sum_i^{120}\frac{[\theta_{th}(z_i)-\theta_{obs}^i]^2}{\sigma^2_i},
\end{equation}
where $\theta_i \pm \sigma_i$ is the observed angular size and its $68\%$ confidence level uncertainty at the redshift $z_i$ and $\theta_{th}$ is the theoretical angular size, which expressed in terms of the angular diameter distance, $D_A(z) = d_L(z)/(1+z)^2$, is
\cite{Sandage:1988}
\begin{equation}
    \theta(z) = \frac{l_m}{D_A(z)}\,,
\end{equation}
where $l_m$ is an intrinsic length  fixed at the value $l_m=11.03 \pm 0.25$ pc found by \cite{ShuoQSO:2017}.

%%%%%%%%%%%%%%%%%%%%%%%%%%%%%%%%%%%%%
\subsection{Hydrogen II Galaxies}
%%%%%%%%%%%%%%%%%%%%%%%%%%%%%%%%%%%%%
A total of 181 distance modulus measurements coming from Hydrogen II galaxies spanning a redshift region $0.01<z<2.6$ are included in the Bayesian analysis \cite{GonzalezMoran2019, Gonzalez-Moran:2021drc}. These HIIG are compact galaxies with low mass with the characteristic that their luminosity is dominated by young massive burst of star formation \cite{Chavez2014}. This allows to extract a correlation between the luminosity ($L$) and 
the inferred velocity dispersion ($\sigma$) of the ionized gas. In this case, the $\chi^2$-function can be written as
\begin{equation}\label{eq:chi2_HIIG}
    \chi^2_{{\rm HIIG}}=\sum_i^{181}\frac{[\mu_{th}(z_i, {\Theta})-\mu_{obs}^i]^2}{\epsilon_i^2}.
\end{equation}
The uncertainties of $\mu_{obs}^i$ at the redshift $z_i$ is $\epsilon_i$, being $\mu_{obs}$ the measured distance modulus obtained through
\begin{equation}
    \mu_{obs} = 2.5(\alpha + \beta\log \sigma -\log f - 40.08)\,,
\end{equation}
where $\alpha$ and $\beta$ are the intercept and slope of the $L$-$\sigma$ relation respectively, and $f$ is the measured flux. 

The theoretical distance modulus is 
\begin{equation}
    \mu_{th}(z, \Theta) = 5 \log_{10} \left [ \frac{d_L(z)}{1\,{\rm Mpc}}\right] + 25,
\end{equation}
where $d_L$ is the luminosity distance measured in Mpc.

%%%%%%%%%%%%%%%%%%%%%%%%%%%%%%%%%%%%%%
\subsection{Baryon Acoustic Oscillations}
%%%%%%%%%%%%%%%%%%%%%%%%%%%%%%%%%%%%%%

Baryon Acoustic Oscillation signal is a powerful tool to constrain cosmological parameters and is produced in the recombination epoch coming from the baryon-photon interactions \cite{Dwyer:2020, Anselmi:2023}. The typical measurements from BAO are the ratio $D_V(z)/r_d$ between the dilation scale $D_V(z)$ measured at the redshift $z$ and the size of the sound horizon at the drag epoch $r_d=r_s(z_d)$, the ratio $D_M(z)/r_d$ where $D_M(z)$ is the comoving angular diameter, and $D_H(z)/r_d = (c/H(z))/r_d$. The dilation scale is defined as \cite{Wigglez:Eisenstein2005}
\begin{equation}
    D_V(z) = [z\,D_H(z)\,D_M^2(z)]^{1/3}\,,
\end{equation}
the  sound horizon at $r_d$ is given by
\begin{equation}
    r_d = \int_{z_d}^\infty \frac{c_s(z)dz}{H(z)}\,,
\end{equation}
where $c_s(z)$ is the sound speed and 
\begin{equation}
    D_M(z) = \frac{c}{H_0} \int_0^z \frac{dz'}{E(z')}\,, \label{DM}
\end{equation}
is expressed for a flat cosmology. 
To confront the cosmological model to these measurements, we built a $\chi^2$ function of the form
\begin{equation}
    \chi^2_{\rm BAO} = \sum_i^N \left(\frac{O^i_{th}(z_i)-O^i_{obs}}{\sigma_i} \right)^2
\end{equation}
where $O_{th}$ is the theoretical estimate of the observed counterpart $O_{obs}^i\pm \sigma_i$ at the redshift $z_i$, and the sum runs over the size of the sample. In this case, we will consider two samples, one comes from Sloan Digital Sky Survey (SDSS) and Dark Energy Survey (DES) and the second from the recent results from Dark Energy Spectroscopy Instrument (DESI). For this work we use $z_d = 1089.80 \pm 0.21$ \cite{Planck:2018}. For more details about this sample, see Table 3 of \cite{Alam_2021}.

\subsubsection{BAO measurements from SDSS and DES}

A total of 15 BAO measurements are used, 14 uncorrelated BAO quantities are reported in \cite{Alam_2021} which cover a region $0.15<z<2.33$, and the recent BAO point $D_M(z_{eff})/r_d = 19.51\pm0.41$ at the effective redshift $z_{eff}=0.85$ reported by DES  \cite{descollaboration2024dark}. We refer to this dataset as SDSS+DES.

\subsubsection{BAO measurements from DESI}

During the first year of DESI observations, the collaboration reported new 12 BAO measurements derived from BAO signatures in galaxy, quasars and Lyman-$\alpha$ forest tracers in the redshift range $0.1<z<4.16$  \cite[see table 1 of][]{desicollaboration2024desi}. 
We use this sample to compare the parameter constraints with the SDSS dataset. We refer to this dataset as DESI-1Y.

%%%%%%%%%%%%%%%%%%%%%%%%%%%%%%%%%%%%%%
\section{Results} \label{Results}
%%%%%%%%%%%%%%%%%%%%%%%%%%%%%%%%%%%%%%

\begin{figure*}
    \centering
    \includegraphics[width=0.6\textwidth]{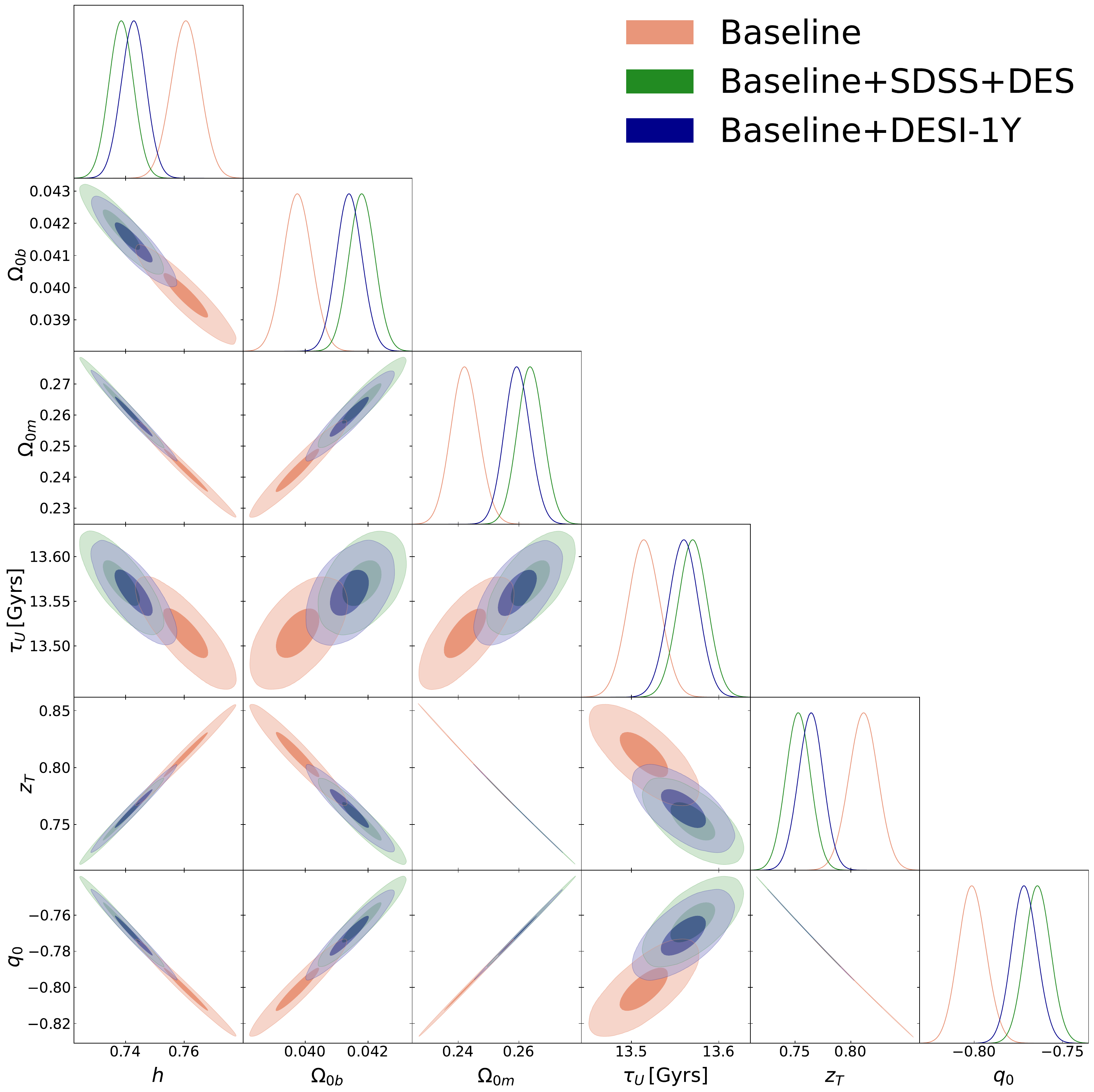}
    \caption{2D parameter distributions at $1\sigma$ (inner region) and $3\sigma$ (outermost region) for PEDE cosmology.}
    \label{fig:contours_baseline}
\end{figure*}

This section is devoted to present our results for the PEDE cosmology. Figure \ref{fig:contours_baseline} shows the  2D contour distribution at 68\% ($1\sigma$) confidence level (CL)  and 99.7\% ($3\sigma$) CL of the parameters, and the 1D posterior distribution using the  baseline dataset (CMB+CC+SNIa+QSO+HIIG), baseline+SDSS+DES, and baseline+DESI-1Y datasets, respectively. Best fit parameters and their uncertainties at $1\sigma$ CL obtained using these datasets and only SDSS+DES and DESI-1Y data are shown in Table \ref{tab:bf_model}. 

\begingroup

\setlength{\tabcolsep}{10pt} % Default value: 6pt
\renewcommand{\arraystretch}{1.5} % Default value: 1

\begin{table*}
	\centering
	\caption{Best fit values and their $1\sigma$ of uncertainties for PEDE cosmology using Baseline (CMB+CC+SNIa+QSO+HIIG), Baseline+SDSS+DES,  Baseline+DESI-1Y. The last two columns are the corresponding values when only SDSS+DES and DESI-1Y are considered respectively. Baseline sample is the sum of DA, Pantheon+, HIIG, QSO datasets.}
	\label{tab:bf_model}
	\begin{tabular}{lccccc} % four columns, alignment for each
    \hline
    Parameter & Baseline & Baseline+SDSS+DES & Baseline+DESI-1Y & SDSS+DES & DESI-1Y \\
    \hline
        \multicolumn{6}{c}{PEDE} \\
 $h$  & 	$0.761^{+0.005}_{-0.005}$  & 	$0.739^{+0.004}_{-0.004}$  & 	$0.743^{+0.004}_{-0.004}$  & 	-  & 	-  \\ [0.9ex] 
 $\Omega_{0b}$  & 	$0.040^{+0.001}_{-0.001}$  & 	$0.042^{+0.0003}_{-0.0003}$  & 	$0.041^{+0.0003}_{-0.0003}$  &$0.034^{+0.012}_{-0.020}$  & 	$0.034^{+0.012}_{-0.020}$  \\ [0.9ex] 
 $\Omega_{0m}$  & 	$0.242^{+0.005}_{-0.004}$  & 	$0.264^{+0.004}_{-0.004}$  & 	$0.259^{+0.004}_{-0.004}$  &$0.299^{+0.019}_{-0.018}$  & 	$0.296^{+0.016}_{-0.015}$  \\ [0.9ex] 
 $\tau_U \,[\rm{Gyrs}]$  & 	$13.514^{+0.018}_{-0.018}$  & 	$13.570^{+0.017}_{-0.017}$  & 	$13.560^{+0.017}_{-0.017}$  & 	-  & 	-  \\ [0.9ex] 
 $z_T $  & 	$0.811^{+0.013}_{-0.013}$  & 	$0.753^{+0.011}_{-0.011}$  & 	$0.764^{+0.011}_{-0.011}$  & 	$0.665^{+0.043}_{-0.043}$  & 	$0.673^{+0.036}_{-0.037}$  \\ [0.9ex] 
 $q_0 $  & 	$-0.801^{+0.008}_{-0.008}$  & 	$-0.764^{+0.007}_{-0.007}$  & 	$-0.771^{+0.007}_{-0.007}$  & 	$-0.703^{+0.033}_{-0.031}$  & 	$-0.708^{+0.028}_{-0.026}$  \\ [0.9ex] 
 $\chi^2$  & $5860.78$  & $5946.11$  & $5922.22$  & $21.09$  & $16.28$  \\ [0.9ex] 
\hline 
        \multicolumn{6}{c}{$\Lambda$CDM} \\
 $h$  & 	$0.728^{+0.004}_{-0.004}$  & 	$0.709^{+0.004}_{-0.004}$  & 	$0.712^{+0.004}_{-0.004}$  & 	-  & 	-  \\ [0.9ex] 
 $\Omega_{0b}$  & 	$0.044^{+0.001}_{-0.001}$  & 	$0.046^{+0.001}_{-0.001}$  & 	$0.046^{+0.001}_{-0.001}$  & 	$0.032^{+0.013}_{-0.019}$  & 	$0.033^{+0.013}_{-0.019}$  \\ [0.9ex] 
 $\Omega_{0m}$  & 	$0.256^{+0.005}_{-0.005}$  & 	$0.277^{+0.004}_{-0.004}$  & 	$0.274^{+0.004}_{-0.004}$  & 	$0.294^{+0.020}_{-0.018}$  & 	$0.297^{+0.017}_{-0.016}$  \\ [0.9ex] 
 $\tau_U \,[\rm{Gyrs}]$  & 	$13.897^{+0.015}_{-0.015}$  & 	$13.929^{+0.015}_{-0.015}$  & 	$13.925^{+0.015}_{-0.015}$  & 	-  & 	-  \\ [0.9ex] 
 $z_T $  & 	$0.774^{+0.012}_{-0.012}$  & 	$0.719^{+0.011}_{-0.011}$  & 	$0.726^{+0.011}_{-0.011}$  & 	$0.678^{+0.043}_{-0.044}$  & 	$0.670^{+0.037}_{-0.038}$  \\ [0.9ex] 
 $q_0 $  & 	$-0.778^{+0.008}_{-0.008}$  & 	$-0.741^{+0.007}_{-0.007}$  & 	$-0.746^{+0.007}_{-0.007}$  & 	$-0.712^{+0.034}_{-0.031}$  & 	$-0.706^{+0.029}_{-0.027}$  \\ [0.9ex] 
$\chi^2$  & $5825.18$  & $5908.01$  & $5894.08$  & $16.04$  & $15.33$  \\ [0.9ex] 
 \hline
 $\Delta$AIC & 35.59 & 38.09 &  28.14 & 5.05 & 0.95 \\ [0.9ex]
\hline
	\end{tabular}
\end{table*}

\endgroup

According to our results, the best fit value of $h$ is deviated $2.65\sigma$ from the one reported by SH0ES ($H_0 = 73.04 \pm 1.04\, {\rm km/s/Mpc}$) \cite{Riess_2022} when the baseline is used, and this distance is reduced to $0.77\sigma$ ($1.13\sigma$) for baseline+SDSS+DES (baseline+DESI-1Y). Thus PEDE cosmology could alleviate the Hubble constant tension without adding more free parameters than those presented in $\Lambda$CDM, which is consistent with those results obtained by \cite{PEDE:2019ApJ, Pan:2019hac, wang2024constraining} using different samples. It is worth to mention that there is a degeneracy between $h$ and $r_d$ when only SDSS+DES data are used, consequently we do not give a value for $h$. With respect to matter density, we find  higher values (around $13\%$ greater) for the single BAO samples than those using the rest of the data.  This result is consistent with those reported in \cite{PEDE:2019ApJ}. 
However, for both SDSS+DES and DESI-1Y samples, we find that all quantities are in agreement for both PEDE and $\Lambda$CDM models (see Tab. \ref{tab:bf_model} ). 
Additionally, we find a negative correlation between $\Omega_{m0}$ and $h$ (or $H_0$) as reported in \cite{desicollaboration2024desi}. Furthermore, the value of $h$ increases when BAO samples are added to our baseline sample, which is in agreement with the shift presented in \cite{desicollaboration2024desi} when DESI-1Y is included to the CMB sample.
Regarding the age of the Universe $\tau_U$, we find that PEDE model predicts a younger Universe ($\sim3\%$) when using baseline, baseline+SDSS+DES, and baseline+DESI-1Y than the ones predicted by $\Lambda$CDM. Notice that, due the degeneracy between $\tau_U$ and $h$,  those values are not reported when only BAO data is used.

\begin{figure*}
    \centering
    \includegraphics[width=0.32\textwidth]{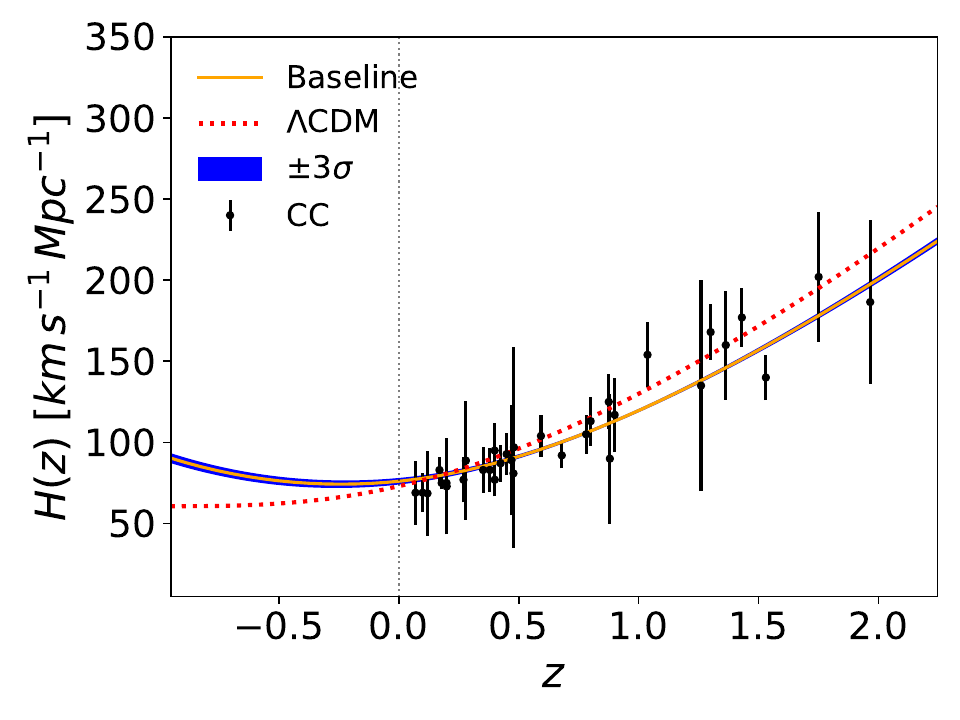}
    \includegraphics[width=0.32\textwidth]{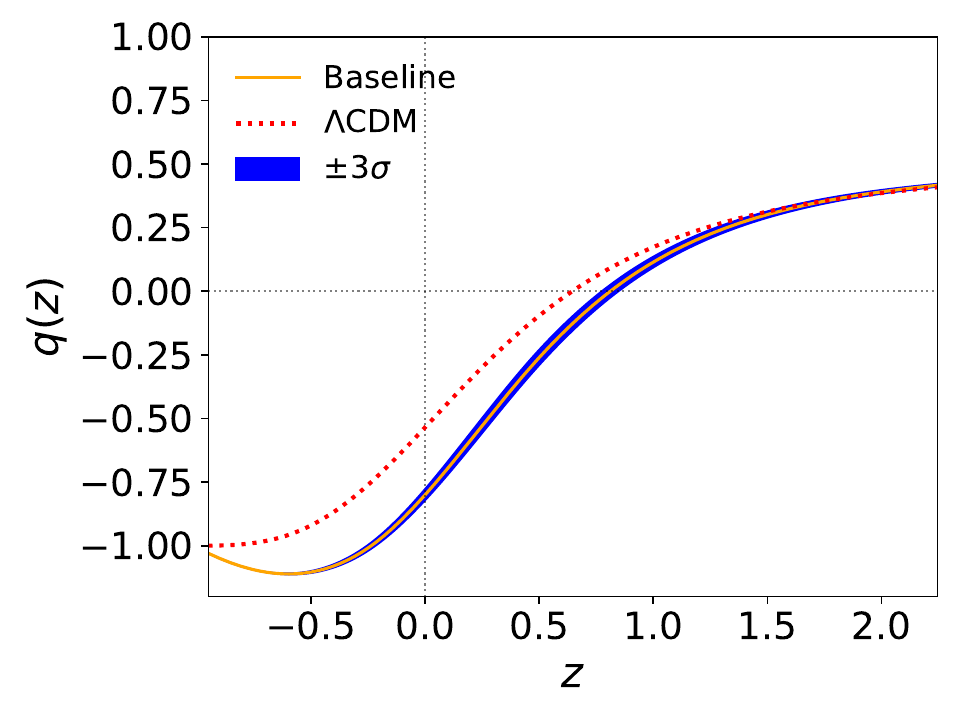}
    \includegraphics[width=0.32\textwidth]{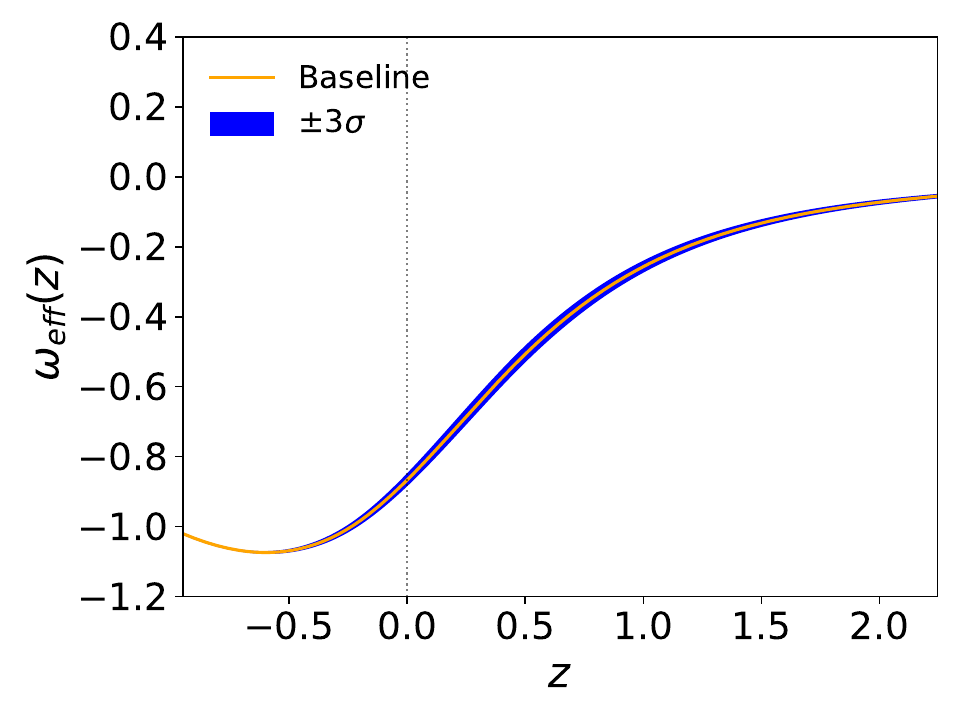}\\
    \includegraphics[width=0.32\textwidth]{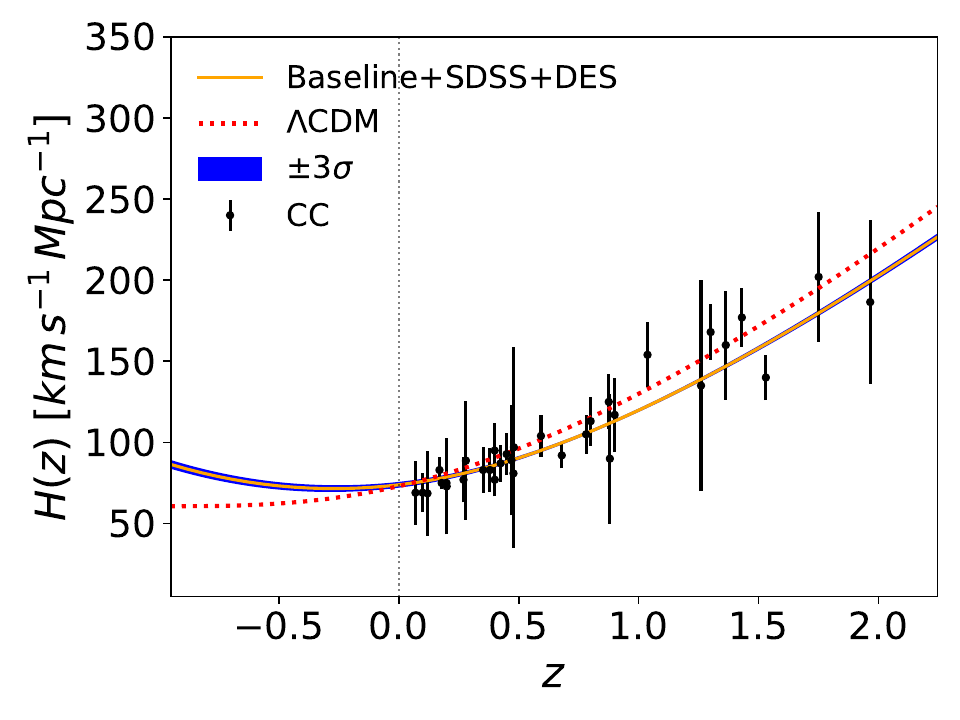}
    \includegraphics[width=0.32\textwidth]{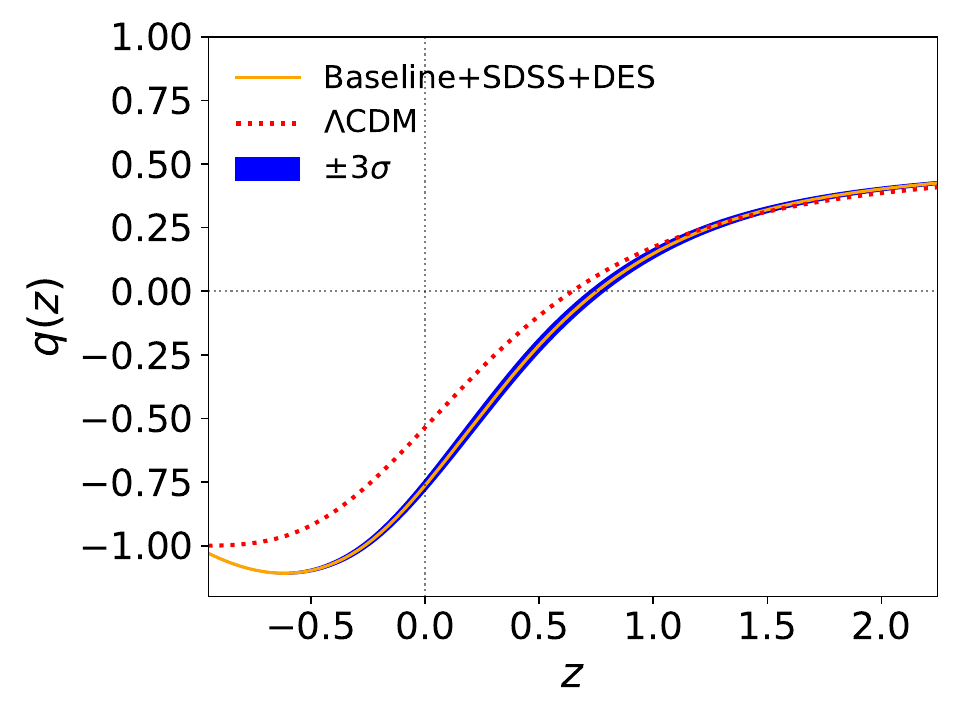}
    \includegraphics[width=0.32\textwidth]{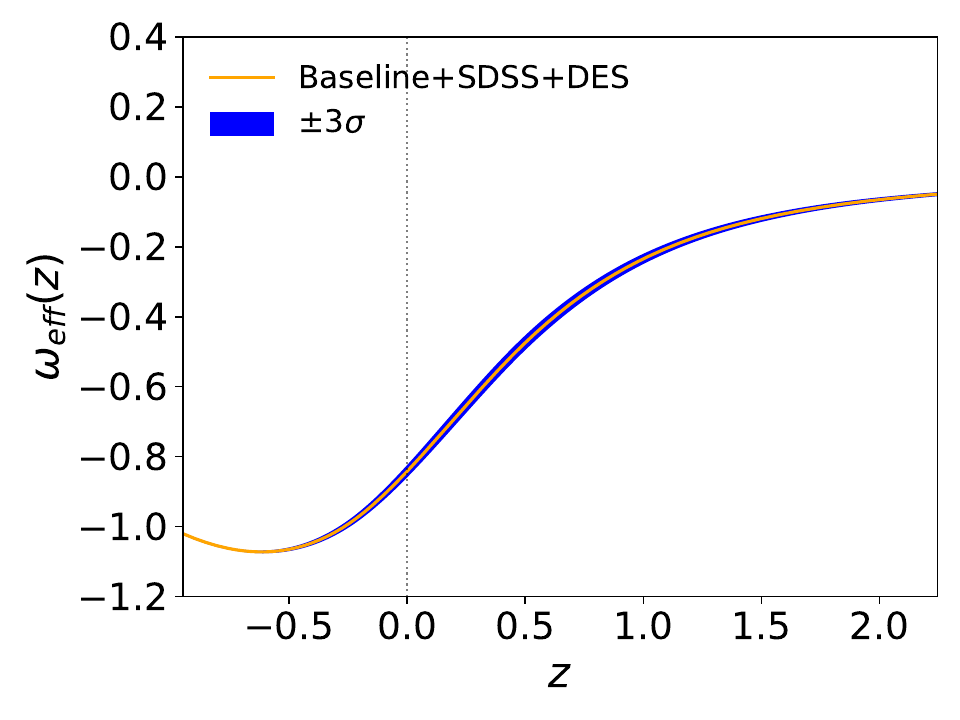}\\ 
        \includegraphics[width=0.32\textwidth]{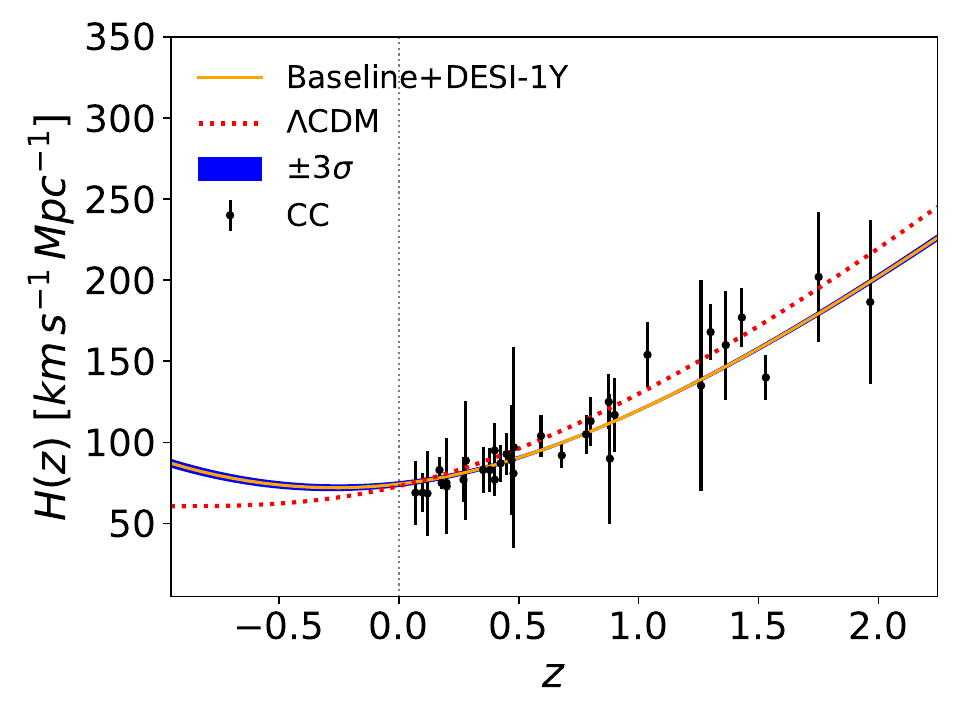}
    \includegraphics[width=0.32\textwidth]{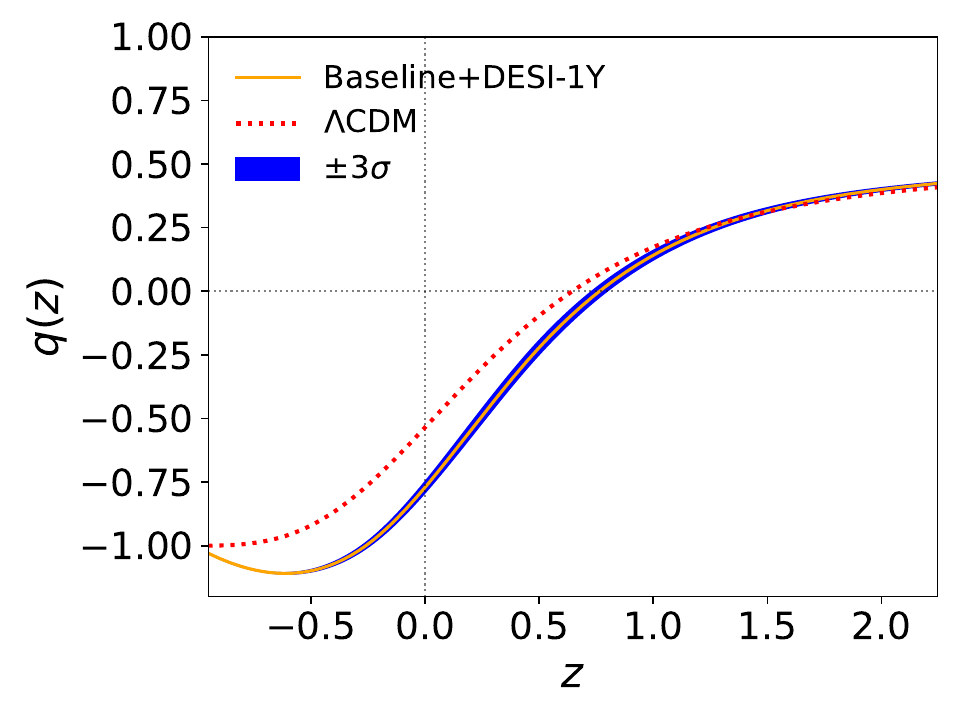}
    \includegraphics[width=0.32\textwidth]{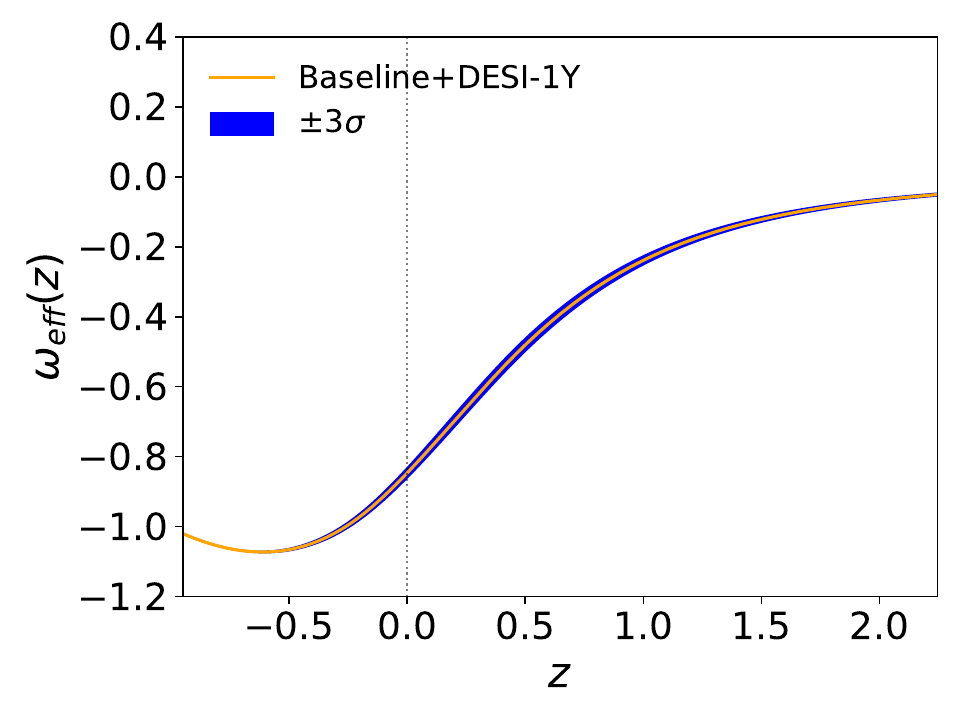}
    \caption{Reconstruction of the Hubble (left panel), the deceleration parameters (middle panel), and the effective equation of state  (right panel) for PEDE model in the redshift range $-0.95<z<2.2$ using baseline (top panel), baseline+SDSS+DES (middle panel) and baseline+DESI-1Y (bottom panel). Blue bands represent $3\sigma$ CL uncertainties of the  PEDE cosmographic parameters. The dotted red line represents the evolution for the standard $\Lambda$CDM model. Baseline refers to the combined sample of CMB+CC+SNIa+QSO+HIIG. CC represents the cosmic chronometers sample described in Sec. \ref{sec:cc}}
    \label{fig:cosmography_pede}
\end{figure*}

Figure \ref{fig:cosmography_pede} displays the Hubble parameter, deceleration parameter and the effective EoS in the redshift range $-0.95 < z < 2.2$ for baseline, baseline+SDSS+DES, and baseline+DESI-1Y datasets respectively. PEDE model predicts lower values of $H(z)$ in the region $z\gtrsim 0.2$ and higher values for the future than those estimated by the standard cosmology, indicating a higher expansion rate in the future. About the deceleration parameter, we extract its current value $q_0$ for each dataset and find that they  deviate more than $3\sigma$ from those estimated from $\Lambda$CDM. Similarly, we obtain the corresponding values for the redshift of the deceleration-acceleration transition $z_T$, for which PEDE cosmology predicts an earlier transition than the standard cosmology. Additionally, $q(z)$ from both cosmologies converges in the future to the value $q(z\to -1)\to-1$. Regarding the $w_{eff}$, we observe that the PEDE cosmology behaves as a fluid that mimics a quintessence fluid in the past and as a phantom fluid in the future $z\lesssim-0.1$. Furthermore, in the region $z\gtrsim 0.7-0.8$, the effective EoS does not allow a Universe acceleration according to the restriction $w<-1/3$ imposed by general relativity, instead the Universe is going to the matter phase in which $w_{eff}\to 0$ as expected. 

As a test, we compare our baseline results with those obtained by adding 14 new HIIG points reported in \cite{Llerena_2023, degraaff2023ionised} and collected in \cite{chavez2024reconstructing} that covers a redshift region $2.41<z<7.43$. Five of them were measured using the James Webb Space Telescope (JWST) as part of the JWST Advanced Deep Extragalactic Survey \cite{bunker2024jades}. We do not find a significant deviation from our baseline results shown in Tab. \ref{tab:bf_model}.

Finally, as statistical comparison between PEDE and $\Lambda$CDM, the Akaike information criterion (AIC)  \cite{AIC:1974, Sugiura:1978} is used and defined as AIC$\equiv\chi^2 + 2k$, where $k$ is the number of free parameters. This criterion considers a preferred model as the one with the lowest AIC value. Although both cosmological models under consideration have the same number of degrees of freedom, we can use the AIC rules to perform the statistical test. In this vein, if a difference in the AIC value between a given model and the best one ($\Delta$AIC) is less than $4$, both models are equally supported by the dataset. For a $\Delta$AIC value in the interval $4<\Delta$AIC$<10$, the data still support the given model but less than the preferred one. For $\Delta$AIC$>10$, the observations do not support the given model. Based on these criteria, we find that  $\Lambda$CDM is preferred for baseline, baseline+SDSS+DES, baseline+DESI-1Y, and SDSS+DES. In contrast, both models are equally supported when DESI-1Y data is used. 
Let us compare these results with those reported in the literature. Shafieloo et. al. \cite{PEDE:2019ApJ} find a value of $\Delta\chi^2= 3.638$ between PEDE and $\Lambda$CDM using CMB Planck 2018, thus both models are comparable statistically. They find 
a difference of $\Delta\chi^2 =-15.332$ where PEDE is preferred when a prior on SH0ES value of $H_0$ is added. Authors \cite{DESI:2024kob} constrain the Generalized Emergent Dark Energy model (notice that $\Delta=1$ is reduced to PEDE) using only DESI-1Y sample and find $\Delta \chi^2 = -0.04$, a difference of $\Delta \chi^2 = -5.7$ using DESI-1Y+CMB and $\Delta\chi^2=-4.8$ using DESI-1Y+CMB+PANTHEONPLUS. It is worth to mention that only for DESI-1Y+CMB the GEDE parameter $\Delta=0.81\pm 0.40$ is consistent within $1\sigma$ with PEDE model. Furthermore, \cite{Pan:2019hac} finds a shift between both models of $\Delta\chi^2 \approx -1.5$ using CMB data, and this increase up to $\Delta \chi^2\approx -19$ for CMB+BAO+PANTHEONPLUS, concluding that PEDE cosmology is strongly preferred.

%%%%%%%%%%%%%%%%%%%%%%%%%%%%%%%%%%%%%%
\section{Summary and Discussions} \label{SD}
%%%%%%%%%%%%%%%%%%%%%%%%%%%%%%%%%%%%%%

PEDE cosmology is a successful model which describes the behavior of the DE component at late times by proposing an emergent DE, i.e. it is not an event that arise from the Big Bang as it happens with the cosmological constant, but instead the DE is described by a dynamical EoS and thus conserving the same degrees of freedom as the standard cosmology. PEDE cosmology has been confronted to several datasets previously, given interesting results regarding the Hubble tension, $S_8$ tension and others \cite{DES:2021wwk,KiDS:2020suj, Heymans_2021} (for excellent reviews of the $ \Lambda$CDM status see \cite{Peebles:2024txt,Perivolaropoulos:2021jda}). For this reason, this work was devoted to revisit this emergent model by confronting it to recent data from SNIa (Pantheon+), QSO, HIIG, BAO (SDSS+DES and DESI), CC, and CMB distance prior. In this sense, the baseline data is defined as the sum of CMB, CC, SNIa, QSO, HIIG. Besides using the baseline, BAO and DESI-1Y databases by themselves, we included the combination of baseline+SDSS+DES and baseline+DESI-1Y.
According to Table \ref{tab:bf_model}, the value of the Hubble constant is greater than the one from SH0ES \cite{Riess_2022} when using baseline data but it moves towards the latter value when BAO measurements are added. Since the BAO measurements depend on the value of $r_d$ and there is a degeneracy between this value and $h$, both the values of $h$ and the age of the Universe, which is correlated with $h$, are not reported when only the BAO and DESI-1Y samples are considered individually. For matter and baryon densities, the results are in agreement with those obtained by the standard model, which is not the case for the redshift of the transition accelerated-decelerated Universe as PEDE cosmology predicts an earlier transition than the standard model. Similarly, the effective EoS has a trend to the $\Lambda$CDM EoS value at the future, and today ($z=0$) the $w_{eff}$ mimics a quintessence DE. Additionally, for the age of the Universe $\tau_U$ we find that PEDE model predicts a younger Universe ($\sim3\%$) when using baseline, baseline+SDSS+DES, and baseline+DESI-1Y than the one predicted by $\Lambda$CDM. 
This result is relevant when compared it with the age of the oldest objects in our Universe, for example, a sub-giant moving off the main sequence called HD 140283 has an estimated age of  $\tau_U=14.5\pm0.8$Gyr obtained through stellar evolutionary models, while for a nuclear reaction of $^{14}N(p,\gamma)^{15}O$ of the CNO burning cycle give us an age of $\tau_{U}=14.2\pm0.6(\pm{\rm systematics})$Gyr (see \cite{Verde:2013wza} and references therein), clearly in tension no only with our results but also with the standard $\Lambda$CDM model \cite{Perivolaropoulos:2021jda}. Furthermore, \cite{Valcin_2020} reports  the average age of the oldest globular clusters as $\tau_{GC}=13.32 \pm 0.10 (\rm stat)\pm 0.50 ({\rm syst})\,$Gyrs  which is about $\sim0.5\sigma$ below our estimates of the Universe age using PEDE cosmology.

Emergent DE models are the cornerstone to understand the current Universe acceleration without the need of a cosmological constant and its unsolvable problems. Mainly because the emergent models do not appear with the Big Bang as cosmological constant does, they manifest only at recent epochs and are produced by some unexplored events.
Although our results based on AIC disfavour PEDE cosmology using the combined samples, PEDE could be the key to alleviate the problem in $H_0$ as Ref. \cite{Pan:2019hac} states, noticing a reduction within $60\%$ confidence level with an improvement of the $\chi^2$ for CMB. The key for this alleviation could be the presence of a dynamic DE at $z\to0$ instead of the cosmological constant present  since the beginning of the Universe (although almost negligible for all the early epochs). However, it is not yet clear which is the dynamical event producing the functional form of PEDE, being the best candidate a scalar field with the capability to reproduce Eq. \eqref{eq:PEDE} as Ref. \cite{Mortonson_2009} argue. 
Finally, it is worth mentioning that we use several samples related to luminous distances from different astrophysical objects in addition to type Ia supernovae, such as quasars, type II hydrogen-dominated galaxies, among others. As happened with SNIa, debates persist over their physical nature and measurements (for instance, the calibration precision). However, those are available samples and their utility can only be better understood by use.

\begin{acknowledgments}
We thank anonymous referees for thoughtful remarks and suggestions. A.H.A. thanks the support from Luis Aguilar, 
Alejandro de Le\'on, Carlos Flores, and Jair Garc\'ia of the Laboratorio 
Nacional de Visualizaci\'on Cient\'ifica Avanzada. M.A.G.-A. acknowledges support from c\'atedra Marcos Moshinsky (MM), Universidad Iberoamericana for support with the SNI grant and the numerical analysis was also carried out by {\it Numerical Integration for Cosmological Theory and Experiments in High-energy Astrophysics} (Nicte Ha) cluster at IBERO University, acquired through c\'atedra MM support. A.H.A, V.M. and M.A.G.-A acknowledge partial support from project ANID Vinculaci\'on Internacional FOVI220144. M.L.M.M. thanks to the SNII-Me\'xico, CONAHCyT for support.
\end{acknowledgments}

\bibliography{main}

\end{document}